\documentclass[12pt,a4paper]{article}
\usepackage{latexsym}
\usepackage{epsf}
\makeatletter
\@addtoreset{equation}{section}
\makeatother

\def\unit{\hbox to 3.3pt{\hskip1.3pt \vrule height 7pt width .4pt \hskip.7pt
\vrule height 7.85pt width .4pt \kern-2.4pt
\hrulefill \kern-3pt
\raise 4pt\hbox{\char'40}}}

\def\x{\times}

\def\be{\begin{equation}}
\def\ee{\end{equation}}
\def\bea{\begin{eqnarray}}
\def\eea{\end{eqnarray}} 
\def\nn{\nonumber \\}

\def\part{\partial}
\def\tfrac#1#2{{\textstyle{#1\over #2}}}

\def\cG{{\cal G}}

\def\cH{{\cal H}}

\def\hi{{\hat \imath}}
\def\hj{{\hat \jmath}}
\def\hk{{\hat k}}
\def\hl{{\hat l}}

\def \alfa {2 \pi \alpha'}

\def \j {{\sf g}} 

\begin{document}
\begin{flushright}
\footnotesize
UG-12/98\\
QMW-PH-98-26 \\
June, $1998$
\normalsize
\end{flushright}

\begin{center}


\vspace{.6cm}
{\LARGE {\bf 5-branes, KK-monopoles and T-duality}}

\vspace{.9cm}

{
{\bf Eduardo Eyras}
\footnote{E-mail address: {\tt E.A.Eyras@phys.rug.nl}},
{\bf Bert Janssen}
\footnote{E-mail address: {\tt B.Janssen@phys.rug.nl}}
\footnote{Address after September 1st: Instituto de F\'{\i}sica
Te\'orica, C-XVI, Universidad Aut\'onoma de Madrid, 28049 Madrid, Spain}\\
{\it Institute for Theoretical Physics\\
University of Groningen\\
Nijenborgh 4, 9747 AG Groningen, The Netherlands}
}

\vspace{.5cm}

{
{\bf Yolanda Lozano}
\footnote{E-mail address: {\tt Y.Lozano@qmw.ac.uk}}\\
{\it Physics Department\\
Queen Mary \& Westfield College\\
Mile End Road, London E1 4NS, U.K.}
}

\vspace{.2cm}


\vspace{.2cm}

\vspace{.8cm}


{\bf Abstract}

\end{center}

\begin{quotation}

\small

We construct the explicit worldvolume effective actions of the type IIB
NS-5-brane and KK-monopole. These objects are obtained through a $T$-duality 
transformation from the IIA KK-monopole and the IIA NS-5-brane
respectively. We show that the worldvolume field content of these
actions is precisely that necessary to describe their worldvolume solitons.
The IIB NS-5-brane effective action is shown to be related to 
the D-5-brane's by an $S$-duality
transformation, suggesting the way to construct 
$(p,q)$ 5-brane multiplets. 
The IIB KK-monopole is described by a gauged sigma model, 
in agreement with the general picture for KK-monopoles, and behaves as a
singlet under $S$-duality. We derive the explicit $T$-duality rules 
NS-5-brane $\leftrightarrow$ KK, which we use for the construction of the
previous actions, as well as NS-5 $\leftrightarrow$ NS-5, and KK 
$\leftrightarrow$ KK.

\end{quotation}

\vspace{1cm}

\newpage

\pagestyle{plain}


\newpage
\section{Introduction}

The worldvolume effective actions that provide the sources for the brane
solutions of type II superstring theories have proved to reveal a great
deal of information about the dynamics of these objects
\cite{Strominger,Townsend1,CM,Gibbons}. The Born-Infeld
(BI) field present in the worldvolume of D-branes describes the flux of a
fundamental string ending on the brane
\cite{Polchinski,D-lectures}. The self-dual 2-form present in the 
action of the M-5-brane (IIA NS-5-brane) describes the dynamics 
of the string-like boundary of an M-$2$-brane (D-$2$-brane) 
on the M-$5$-brane (NS-$5$-brane) \cite{Strominger}.
In general, fields propagating on a brane worldvolume
describe the dynamics of the boundaries of other branes
ending on it. These configurations can be obtained
from configurations of intersecting closed branes where
one of the branes opens up to lean its boundary on the other brane
\cite{Townsend1,Ricc}. These intersection regions, as well as the boundaries
of branes ending on another branes, are described by worldvolume 
solitons that preserve 1/4 of the supersymmetry of the bulk. From the 
point of view of the particular brane they
preserve 1/2 of its supersymmetry. Therefore the study of the
worldvolume supersymmetry algebra of the brane can be used to reveal its
possible soliton solutions \cite{HLW,BGT}.  

In this paper we construct the worldvolume effective actions 
of the type IIB NS-5-brane and KK-monopole.
Six dimensional gauge theories
obtained as the weak coupling limit of a system of parallel type II
fivebranes have received a lot of attention recently
\cite{Seiberg,Witten1}.
In this limit, interactions on the fivebranes lead to interacting six
dimensional gauge theories, and different gauge groups and supersymmetries
can be obtained depending on the particular choice of fivebranes that is
taken. 

The worldvolume field content that we find in the actions 
for the IIB NS-5-brane and KK-monopole
agrees with
the field content anticipated in \cite{Callan,Hull}, based on 
general arguments
following from the representation theory of six dimensional $N=2$
supersymmetry and $T$-duality. Based on these analysis the
field content of the effective actions could be deduced, even though the
explicit couplings and therefore the explicit effective actions were not
known.

A basic tool in our construction is the $T$-duality symmetry between 
the type IIA and type IIB
superstring theories \cite{BHO}. For D-branes $T$-duality relates the
direct dimensional reduction of a given D-$p$-brane with the double dimensional
reduction of a D-$(p+1)$-brane in the dual theory 
\cite{D-lectures,Eric-Mees,Green-Hull-Townsend}.
For other solitonic objects, like
the NS-5-branes and the KK-monopoles considered in this paper, $T$-duality
can work in two different ways, depending on whether we are dualizing with
respect to a worldvolume coordinate or with respect to a transverse
coordinate. The transformations of the target space fields are of course
the same but the way the worldvolume fields transform differ from one case
to the other. In this paper we will analyze both
possibilities. 

The IIB NS-5-brane is related by
$T$-duality along a transverse coordinate 
with the IIA KK-monopole. Therefore, its worldvolume theory must
be described by the six dimensional $(1,1)$ vector supermultiplet, which
contains 4 scalars and one vector. The IIB KK-monopole is related by
$T$-duality (also along a transverse coordinate, in particular its Taub-NUT 
direction) with the IIA NS-5-brane, which implies that it must be described
by a $(2,0)$ vector supermultiplet, which contains a selfdual 2-form and 5
scalars. 

The D-5-brane and
the NS-5-brane solutions of type IIB supergravity are related by an
$S$-duality transformation. Therefore, it is natural to expect that 
the corresponding 
effective actions must be related by the same kind of transformation. 
However, starting from the D-5-brane effective action it is not clear how
the worldvolume fields should transform under $S$-duality, in particular
whether a 
worldvolume duality transformation needs to be done, as happens with the
relation between F- and D- strings \cite{D2-brane}.
Our approach in this paper is to construct the action of the IIB NS-5-brane
starting from the action of the IIA KK-monopole
\cite{BEL}. At the level of the supergravity solutions one can perform a
$T$-duality transformation along the isometry direction of the Taub-NUT space 
of the monopole in order to obtain the 5-brane supergravity solution. 
At the level of the worldvolume effective actions we proceed in the same
way.
Then we check that the resulting action is indeed $S$-dual to the D-5-brane
and derive the $S$-duality transformation rules of the corresponding
worldvolume fields. An interesting feature is that a worldvolume Poincar{\'e} 
duality is not needed in this case.

On the other hand, $T$-duality also relates both type II NS-$5$-branes 
(KK-monopoles).
We explicitly work out these duality transformations. They provide a check
for the action of the IIB NS-5-brane (IIB KK-monopole) 
constructed from the IIA KK-monopole (IIA NS-5-brane). 

NS-5-branes and KK-monopoles couple to dual space time potentials, 
for which no explicit
$T$-duality transformation rules have been derived. 
We fill this gap and derive as well the explicit 
$T$-duality rules for the new background fields that couple to the 
KK-monopole effective actions.

The structure of the paper is as follows.
In section 2 we derive the action of the IIB NS-5-brane. 
We then check that the resulting
action is related to the D-5-brane effective action by an $S$-duality
transformation. 

In section 3 we construct the action of the IIB KK-monopole. 
In this case we perform a $T$-duality transformation in the worldvolume
action of the IIA NS-5-brane \cite{BLO}. The resulting action is a singlet
under $S$-duality, a property that can also be used to derive
the field content of the IIB KK-monopole \cite{Hull}.

In section 4 we present the worldvolume $T$-duality rules that map the 
NS-5-branes of the type IIA and IIB theories onto each other. 
In this case a further worldvolume duality transformation is required in
order to show the equivalence between the two actions.
In section 5
we do the same for the KK-monopole effective actions.
These T-dualities establish a map between $(1,1)$ and $(2,0)$ six
dimensional supersymmetric theories \cite{Callan}. The analysis of the
fermionic parts should reveal the reversing of the space-time chirality
under the T-duality transformation.

Appendix A contains the $T$-duality rules for the dual background potentials
coupled to the NS-5-branes and KK-monopoles, as well as the transformations 
of the new target space fields present in the KK-monopole effective
actions.
Appendices B and C summarize the gauge transformation rules of the target
space and worldvolume fields considered in the paper.

Finally in section 6 we present our conclusions and open problems.

\vspace{1cm}  

We finish the introduction by summarizing the target space fields 
occurring in the IIA and
IIB superstring theories in order to set up the notation for the rest of
the paper. They can be found in tables 1 and 2.

\begin{table}[!ht]
\begin{center}
\begin{tabular}{||c|c||c|c||}
\hline\hline
Target space & Gauge     & Dual  & Gauge    \\
Field        & Parameter & Field & Parameter \\
\hline\hline
${g}_{{\mu}{\nu}}$, $\phi$
& $-$ & $-$ & $-$ \\
\hline
$B_{\mu \nu}$ & $\Lambda_\mu$ & ${\tilde B}_{\mu_1 \dots \mu_6}$ 
& ${\tilde \Lambda}_{\mu_1 \dots \mu_5}$ \\
\hline
$C^{(1)}_\mu$ & $\Lambda^{(0)}$ & $C^{(7)}_{\mu_1 \dots \mu_7}$ 
& $\Lambda^{(6)}_{\mu_1 \dots \mu_6}$ \\ 
\hline
${C}^{(3)}_{{\mu}{\nu}{\rho}}$ & $\Lambda^{(2)}_{\mu \nu}$ & 
$C^{(5)}_{\mu_1 \dots \mu_5}$ 
& $\Lambda^{(4)}_{\mu_1 \dots \mu_4}$ \\
\hline
$k_{\mu}$ & $-$ & ${N}_{\mu_1 \dots \mu_7}$ & 
${\Sigma}^{(6)}_{\mu_1\dots\mu_6}$ \\
\hline\hline
\end{tabular}
\end{center}  
\caption{\label{table1} {\bf Target space fields of the type IIA 
superstring}.
 \footnotesize
The type IIA background contains the NS-NS sector:
$( g_{\mu \nu}, \phi, B_{\mu \nu})$,
the RR sector:
$( C^{(1)}, C^{(3)})$, 
and the Poincar{\'e} duals of the RR fields and the NS-NS 2-form $B$:
$( C^{(5)}, C^{(7)}, {\tilde B})$. The Kaluza-Klein monopole couples to a
new field $N$, dual to the Killing vector associated to the Taub-NUT
isometry, considered as a 1-form $k_\mu$.}
\end{table}

\begin{table}[!ht]
\begin{center}
\begin{tabular}{||c|c||c|c||}
\hline\hline
Target space & Gauge     & Dual  & Gauge    \\
Field        & Parameter & Field & Parameter \\
\hline\hline
${\j}_{{\mu}{\nu}}$, $\varphi$
& $-$ & $-$ & $-$ \\
\hline
${\cal B}_{\mu \nu}$ & $\Lambda_\mu$ & ${\tilde {\cal B}}_{\mu_1 \dots \mu_6}$
 & ${\tilde \Lambda}_{\mu_1 \dots \mu_5}$ \\
\hline
$C^{(0)}$ & $-$ & $-$ & $-$ \\ 
\hline
${C}^{(2)}_{{\mu}{\nu}}$ & $\Lambda^{(1)}_\mu$ 
& $C^{(6)}_{\mu_1 \dots \mu_6}$ 
& $\Lambda^{(5)}_{\mu_1 \dots \mu_5}$ \\
\hline
$k_\mu$ & $-$ & ${\cal N}_{\mu_1\dots\mu_7}$ & 
${\tilde \Sigma}^{(6)}_{\mu_1\dots\mu_6}$ \\
\hline\hline
\end{tabular}
\end{center}  
\caption{\label{table2} {\bf Target space fields of the type IIB 
superstring}.
 \footnotesize
The Type IIB background contains 
the common sector:
$( {\j}_{\mu \nu}, \varphi, {\cal B}_{\mu \nu} )$,
the RR sector: $(C^{(0)}, C^{(2)}, C^{(4)})$, 
and the Poincar{\'e} duals of the 2-forms $C^{(2)}$ and ${\cal B}$:  
$( C^{(6)}, {\tilde {\cal B}})$. We include as well the Poincar{\'e}
dual of the Killing vector $k$, that we denote by ${\cal N}$.
}
\end{table}


\section{The IIB NS-$5$-brane}
\label{IIB-NS-5-brane}


In this section we construct the action of the IIB NS-5-brane, by means 
of a $T$-duality transformation in the effective action of the IIA
KK-monopole. This duality is performed along the Taub-NUT direction
of the monopole.
The action of the IIA KK-monopole was constructed in 
\cite{BEL} and we summarize it in the next subsection.
An interesting feature in this $T$-duality transformation, which makes it
different from the more extensively studied D-brane duality
\cite{D-lectures,Eric-Mees,Green-Hull-Townsend},
is that the number of worldvolume dimensions is the same in 
the original and dual theories\footnote{
This duality was studied  in \cite{Bert} for the heterotic case.}. 
The way it works in this case is explained in detail in subsection 2.2.

\vskip 12pt
\begin{figure}[!ht]
\begin{center}
\leavevmode
\epsfxsize= 10cm
\epsffile{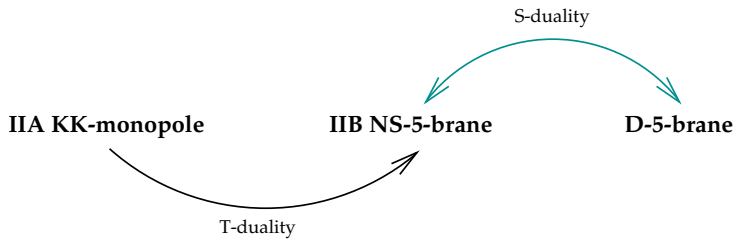}
\caption{\footnotesize
In this figure we depict the way we have derived the IIB NS-5-brane action.
Applying $T$-duality in the IIA KK-monopole action 
along the Taub-NUT direction the type IIB NS-$5$-brane
action is obtained, and it is $S$-dual to the D-$5$-brane effective action.}
\label{fig:cuadro2}
\end{center}
\end{figure}

\subsection{The IIA KK action}

Let us first recall the action of the IIA KK-monopole constructed in
\cite{BEL}. 

The KK-monopole \cite{monopolo} in $D$ dimensions can be considered 
an extended object with
$D-5$ spatial dimensions and one extra isometry direction
transverse to the worldvolume. 
In order to get the right counting of degrees of freedom this isometry
is gauged, such that the effective number of embedding scalars is 3,
fitting in a $D-4$ dimensional vector multiplet. The effective action of a
KK-monopole in $D$ dimensions is then described by a $D-4$ dimensional
gauged sigma model \cite{BJO,Hull-Spence,BEL}\footnote{For some recent 
work on KK-monopoles see \cite{maskk}.}.

\noindent In particular, the action of the (massless) IIA KK-monopole 
is given by \cite{BEL}:
\begin{equation}
\label{m0accion}
\begin{array}{rcl}
S& =&
-T_{{\rm AKK}} \int d^6 \xi \,\,
e^{-2\phi} k^2 \sqrt{1 + e^{2 \phi}k^{-2} (i_k C^{(1)})^2} \times \\
& & \\
& &\hspace{-1.5cm}
\times\sqrt{|{\rm det}(D_iX^\mu D_jX^\nu g_{\mu \nu}
- (\alfa)^2 k^{-2} {\cal K}^{(1)}_i {\cal K}^{(1)}_j +
 { (\alfa)  k^{-1} e^\phi \over \sqrt{ 1 + e^{2 \phi}k^{-2} 
(i_k C^{(1)})^2}}
{\cal K}^{(2)}_{ij})|}\\
& & \\
& & \hspace{-1.5cm}
+ \,\, {1 \over 6!}(\alfa) T_{{\rm AKK}} \int d^6\xi
\,\,  \epsilon^{i_1 \dots i_6}
{\cal K}^{(6)}_{i_1 \dots i_6} \, .\\
\end{array}
\end{equation}
\noindent The covariant derivative is defined by
\begin{equation}
D_i X^\mu = \partial_i  X^\mu + A_i k^\mu \, ,
\end{equation}
where $k^\mu$ is the Killing vector associated to the transverse
target space isometry, and the gauge field $A$ is
a dependent field given by:

\begin{equation}
A_i = k^{-2}\partial_i X^\mu k_\mu \, ,
\end{equation}
with $k^2= - k^\mu k^\nu g_{\mu \nu}$. With this choice the metric is
effectively nine dimensional since the coordinate adapted to the isometry
drops out of the action \cite{BJO}.
 
${\cal K}^{(2)}$ and ${\cal K}^{(1)}$
are the curvatures of the 1- and 0-forms $\omega^{(1)}$ and 
$\omega^{(0)}$, respectively\footnote{
We omit the worldvolume indices with the understanding that they are
completely antisymmetrized. We use this
notation throughout the paper.}:
\begin{equation}
\begin{array}{rcl}
{\cal K}^{(2)} &=& 2 \partial \omega^{(1)} + {1 \over \alfa} (i_k C^{(3)})
-2 {\cal K}^{(1)} (DXC^{(1)}) \, ,\\
& &\\& &\\
{\cal K}^{(1)} &=& \partial \omega^{(0)} - {1 \over \alfa} (i_k B) 
\, .\\
\end{array}
\end{equation}
Finally, ${\cal K}^{(6)}$ is the
WZ curvature of the monopole:
\begin{equation}
\begin{array}{rcl}
{\cal K}^{(6)} &=& \left\{ 6 \partial \omega^{(5)} 
+\frac{1}{2\pi\alpha^\prime}(i_k N)-
30 (i_k C^{(5)})\partial\omega^{(1)}-\frac{15}{2\pi\alpha^\prime}
(i_k C^{(5)})(i_k C^{(3)})\right.
\\ & & \\ & &
-6(i_k {\tilde B}){\cal K}^{(1)}
 -120 (\alfa) DX^{{\mu}} DX^{{\nu}} DX^{{\rho}} 
C^{(3)}_{{\mu} {\nu} {\rho}} 
\, {\cal K}^{(1)} \partial \omega^{(1)}
\\ & & \\ & & 
+{30 \over 2\pi\alpha^{\prime}} DX^{\mu}DX^{\nu} B_{\mu \nu}
 (i_k C^{(3)}) (i_k C^{(3)})
\\ & & \\ & & 
+ {50 \over 2\pi\alpha^{\prime}} DX^{\mu} DX^{\nu} DX^{\rho} 
C^{(3)}_{\mu \nu \rho} 
(i_k B) (i_k C^{(3)})
\\ & & \\ & & 
-30DX^{\mu} DX^{\nu} DX^{\rho} C^{(3)}_{\mu \nu \rho} 
( i_k C^{(3)}) \partial \omega^{(0)}
\\ & & \\ & & 
-180 (\alfa) DX^{\mu} DX^{{\nu}} B_{{\mu} {\nu}} 
\partial \omega^{(1)}\partial \omega^{(1)}
\\ & & \\ & & 
-360 (\alfa)^2 A \partial \omega^{(1)} \partial \omega^{(1)} 
\partial \omega^{(0)}
\\ & & \\ & & 
\left. + 15 (\alfa)^2 { e^{2 \phi} k^{-2} (i_k C^{(1)}) \over 
1 + e^{2 \phi} k^{-2} (i_k C^{(1)})^2}
{\cal K}^{(2)} {\cal K}^{(2)} {\cal K}^{(2)} \right\} \, ,\\
\end{array}
\end{equation}

\noindent where $(i_k L)$ denotes the interior product of 
the field $L$ with the Killing 
vector\footnote{In our notation: $(i_k L)_{\mu_1\dots\mu_r}=
k^{\mu_{r+1}}L_{\mu_1\dots\mu_{r+1}}$.}.
The target space fields $N$ and ${\tilde B}$ are the duals of the
Killing vector, considered as a 1-form $k_{\mu}$, and the NS-NS 2-form $B$,
respectively. 

The worldvolume field content is summarized in Table 3 and the gauge
transformations can be found in Appendix C.1.

\begin{table}[t]
\begin{center}
\begin{tabular}{||c|c|c||}
\hline\hline
Worldvolume     & Field         & $\sharp$ of   \\      
Field           & Strength      & d.o.f         \\
\hline\hline
${X}^{{\mu}}$ & & $10 - 6 - (1) = 3$ \\
\hline
$\omega^{(0)}$ &
 ${{\cal K}}^{(1)}_i$& $1$\\
\hline
$\omega^{(1)}_i$& ${\cal K}^{(2)}_{ij}$& $4$\\
\hline
$\omega^{(5)}_{i_1 \dots i_5}$& ${\cal K}^{(6)}_{i_1 \dots i_6}$ & $-$\\ 
\hline\hline
\end{tabular}
\end{center}  
\caption{\label{table-IIAKK} 
{\bf Worldvolume field content of the IIA KK-monopole}.
{\footnotesize
The field content of the IIA KK-monopole consists on a
1-form ${\omega}^{(1)}$, a 0-form $\omega^{(0)}$
and 3 embedding scalars $X^\mu$, fitting in a 6 dimensional vector
multiplet. One extra 
degree of freedom has been eliminated through the gauging construction. 
The 5-form
$\omega^{(5)}$ describes the tension of the monopole.
}}
\end{table}

Worldvolume fields propagating in the effective action of a given brane
have an interpretation in terms of soliton solutions on the brane. In the
IIA KK-monopole worldvolume action there are four soliton solutions
preserving 1/4 of the supersymmetry of the bulk \cite{Papadopoulos}.
These are a 1-brane, a 2-brane, a 3-brane and a 4-brane solitons. 
The worldvolume
fields present in the KK-monopole effective action are precisely those
describing these soliton solutions.

The worldvolume field $\omega^{(0)}$ 
couples to the 3-brane soliton (through its dual 4-form) 
and $\omega^{(1)}$ couples to the
 0- and the 2-brane solitons (to the latter through its dual 3-form).
Some of the corresponding
intersections are given by:  
$(3|NS5,KK)$, $(0|D2,KK)$ and $(2|D4,KK)$,
where in all these configurations one of the worldvolume directions of 
the brane is wrapped around the 
Taub-NUT direction of the monopole.

The intersections: 
$(3|KK,KK)_1$, $(3|KK,KK)_2$,  
where either both  $S^1$'s coincide or one
worldvolume direction of a monopole is wrapped around the other's
$S^1$, are also possible. These 3-brane solitons couple to the 4-forms dual 
to the embedding scalars, and can
be obtained from the intersections
of two monopoles in M-theory
\cite{Groningen-Boys} by reducing along a common worldvolume direction.

The worldvolume field 
$\omega^{(5)}$ describes the tension of the monopole and
couples to the 4-brane soliton, which is a domain wall in the 
6 dimensional worldvolume as can be seen from the 
intersection $(4|D4,KK)$ \cite{Groningen-Boys}.

\subsection{$T$-duality}

Our aim now is to perform a $T$-duality transformation in this action in order
to obtain the IIB NS-5-brane. The way $T$-duality works in this case
is as follows:
One first reduces the KK-monopole
action along its isometry direction, given by the Killing vector $k$.
The result is mapped into the
direct dimensional reduction of the IIB NS-$5$-brane along one transversal 
coordinate $Z$, which is T-dual to the
worldvolume scalar 
$\omega^{(0)}$ present in the KK-monopole action:

\begin{equation}
\omega^{(0) \prime} = {1 \over \alfa} Z \, .
\end{equation}
   
The KK-monopole and the $5$-brane couple to target space fields for which
the $T$-duality rules have not been worked out in the literature.
In particular, the IIA KK-monopole
couples to the 6-form $i_k N$ and
to $i_k {\tilde B}$, and the IIB NS-5-brane
couples to the 6-form ${\tilde {\cal B}}$, dual to the NS-NS 2-form ${\cal B}$.
We have worked out the explicit $T$-duality rules for these fields. They can
be found in appendix A.

\noindent The  $T$-duality rules for $\omega^{(1)}$ and $\omega^{(5)}$ 
are given by:
\begin{equation}
\begin{array}{rcl}
\omega^{(1) \prime} &=& - c^{(1)} \, ,\\
& &\\
\omega^{(5) \prime} &=& {\tilde c}^{(5)} - 60 (\alfa) Z' 
\partial c^{(1)} \partial c^{(1)} \partial Z \, .\\
\end{array}
\end{equation}
In the last expression $Z'$ is the adapted coordinate to the 
Killing isometry of the monopole, and
${\tilde c}^{(5)}$ is the worldvolume gauge potential that describes the 
tension of the IIB NS-$5$-brane. The last identification is crucial in
order to 
eliminate completely the $Z'$ coordinate associated to the 
isometry of the monopole. Given that it is not a degree of freedom in the
KK-monopole it should not transmit any degree of freedom
to the $5$-brane.

Considering as well the transformation of the target space field 
$(i_k B)$, which can be found in the appendix, we find:

\begin{equation}
\begin{array}{rcl}
{\cal K}^{(2) \prime}_{ij} &=& - {\tilde {\cal F}}_{ij} \, ,\\
\end{array}
\end{equation}
where ${\tilde {\cal F}}$ is the field strength of the worldvolume field
$c^{(1)}$, describing the tension of a D-string \cite{Townsend}:

\begin{equation}
{\tilde {\cal F}}= 2 \partial c^{(1)} + {1 \over \alfa}C^{(2)} \, .
\label{tilde-cal-F}
\end{equation}

Also, under $T$-duality the tensions of the two branes must be identified:
$T_{{\rm AKK}}^{\prime} = T_{{\rm B5}}$.

\subsection{The action of the IIB NS-$5$-brane}

Applying the $T$-duality as described above, we obtain the following action 
for the IIB NS-$5$-brane:

\begin{equation}
\label{5brane-action}
\begin{array}{rcl}
S &=&
-T_{{\rm B5}} \int d^6 \xi \,\,
e^{-2\varphi} \sqrt{1 + e^{2 \varphi} (C^{(0)})^2}
\sqrt{|{\rm det} 
\left( {\j} - (\alfa) {e^{\varphi} \over \sqrt{1 + e^{2 \varphi}
(C^{(0)})^2}} {\tilde {\cal F}} \right)|} \, \\
& & \\
& &+ \,\, {1 \over 6!}(\alfa) T_{{\rm B5}} \int d^6\xi
\,\,  \epsilon^{i_1 \dots i_6}
{\tilde {\cal G}}^{(6)}_{i_1 \dots i_6} \, .\\
\end{array}
\end{equation}
${\tilde {\cal F}}$ is 
defined above in (\ref{tilde-cal-F}) and 
${\tilde {\cal G}}^{(6)}$ is the curvature of the
5-form ${\tilde c}^{(5)}$ describing the tension of the 5-brane.
This 6-form is the gauge invariant Wess-Zumino term:
\begin{equation}
\begin{array}{rcl}
{\tilde {\cal G}}^{(6)} &=& \left\{
6\partial {\tilde c}^{(5)} - {1 \over \alfa}{\tilde {\cal B}}
- {45 \over 2(\alfa)} {\cal B}C^{(2)}C^{(2)} -15 C^{(4)} {\tilde {\cal F}} 
\right. \\ & &\\
& & - 180 (\alfa) {\cal B} \partial c^{(1)}\partial c^{(1)}
-90 {\cal B} C^{(2)} \partial c^{(1)}
\\& &\\& &
\left.
+15 (\alfa)^2 {C^{(0)} \over e^{-2 \varphi} + (C^{(0)})^2} 
{\tilde {\cal F}}{\tilde {\cal F}}{\tilde {\cal F}}
\right\} \, .\\
\end{array}
\end{equation}
The worldvolume field content is summarized in Table 4.
It consists on a vector and 4 scalars, which is the field content of the
six dimensional $(1,1)$ vector supermultiplet \cite{Callan}.
The corresponding gauge transformations can be found in Appendix
C.2.

\begin{table}[t]
\begin{center}
\begin{tabular}{||c|c|c||}
\hline\hline
Worldvolume     & Field         & $\sharp$ of   \\      
Field           & Strength      & d.o.f         \\
\hline\hline
${X}^{{\mu}}$ & & $4$ \\
\hline
$c^{(1)}_i$ & ${\tilde {\cal F}}_{ij}$ & $4$ \\
\hline
${\tilde c}^{(5)}_{i_1 \dots i_5}$
& ${\tilde {\cal G}}^{(6)}_{i_1 \dots i_6}$ & $-$ \\
\hline\hline
\end{tabular}
\end{center}  
\caption{\label{table-IIBNS5} 
{\bf Worldvolume field content  of the IIB NS-5-brane}.
\footnotesize In the case of the IIB NS-$5$-brane,
there are four embedding coordinates $X^\mu$
and one vector field $c^{(1)}$. This field
is the $S$-dual of the BI field $b$, and describes
the flux of a $D$-string ending on the NS-$5$-brane.
The 5-form ${\tilde c}^{(5)}$ describes the tension
of the 5-brane.}
\end{table}

The worldvolume vector field $c^{(1)}$ is associated to the tension of
a D-string, therefore it describes the flux of such an object ending on the
NS-$5$-brane \cite{Strominger,Townsend1}. 
This is the $S$-dual picture of a fundamental string ending on a D-5-brane
(see Figure 2).

\vskip 12pt
\begin{figure}[t]
\begin{center}
\leavevmode
\epsfxsize=10cm
\epsffile{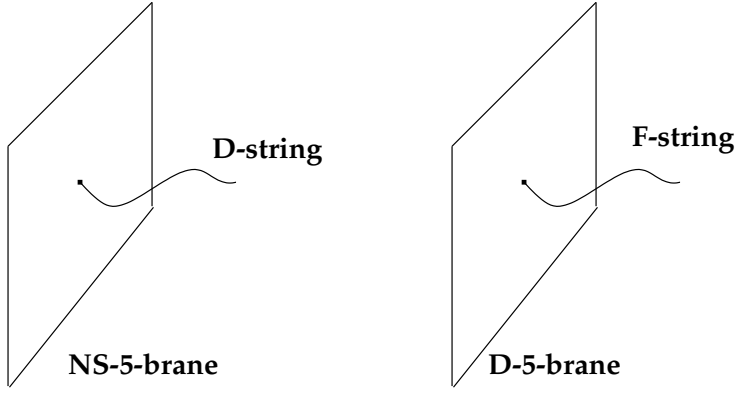}
\caption{\footnotesize
The IIB NS-5-brane worldvolume theory contains a vector $c^{(1)}$,
which is the $S$-dual of the BI field $b$ and
describes the flux of a D-string ending on the 5-brane. This is
the $S$-dual picture of a fundamental string ending on a D-5-brane.}
\label{fig:5brane-duals}
\end{center}
\end{figure}

The action (\ref{5brane-action}) is in fact $S$-dual to the 
D-$5$-brane action.
The following $S$-duality transformation (in the Einstein frame)
\begin{equation}
\begin{array}{rcl}
C^{(0)} &\rightarrow& {- C^{(0)} \over (C^{(0)})^2 + e^{-2 \varphi}} \, ,\\
& &\\
C^{(2)} &\rightarrow& {\cal B} \, ,\\
& &\\
C^{(6)} &\rightarrow& {\tilde {\cal B}}\, ,\\
& &\\
& &\hspace{1.5cm}C^{(4)} \rightarrow C^{(4)}\, ,\\
\end{array}
\begin{array}{rcl}
e^{-\varphi} &\rightarrow& { e^{- \varphi} \over (C^{(0)})^2 + e^{-2 \varphi}}
\, ,\\
& &\\
{\cal B} &\rightarrow& - C^{(2)} \, ,\\
& &\\
{\tilde {\cal B}} &\rightarrow& - C^{(6)} \, ,\\
& &\\& &\\
\end{array}
\end{equation}

\noindent 
in the D-5-brane action:
\begin{equation}
\label{D5brane-action}
S=-T_{D5}\int d^6 \xi\ \Bigl\{ e^{-\varphi}\sqrt{|{\rm
    det}(\j+(2\pi\alpha^\prime){\cal F})|}+
\frac{2\pi\alpha^\prime}{6!}\epsilon^{i_1\dots i_6}
{\cal G}^{(6)}_{i_1\dots i_6}\Bigr\}\, ,
\end{equation}
\noindent 
gives the action (\ref{5brane-action}) describing the
NS-5-brane. 

In (\ref{D5brane-action}) ${\cal G}^{(6)}$ is the WZ curvature:

\begin{equation}
\begin{array}{rcl}
{{\cal G}}^{(6)} &=& \left\{
6\partial {c}^{(5)} + {1 \over \alfa}C^{(6)}
+ {45 \over 2(\alfa)} C^{(2)}{\cal B}{\cal B} -15 C^{(4)} {\cal F} 
\right. 
\\ & &\\& & 
+ 180 (\alfa) C^{(2)} \partial b \partial b
+ 90 C^{(2)}{\cal B} \partial b
\\& &\\& &
\left.-15 (\alfa)^2 C^{(0)} {\cal F}{\cal F}{\cal F}
\right\} \, ,\\
\end{array}
\end{equation}

\noindent ${\cal F}=2 \partial b + {1 \over \alfa}{\cal B}$ and
$b$ is the Born-Infeld field.

The world-volume fields must transform as $SL(2,Z)$ doublets:

\begin{equation}
c^{(1)} \rightarrow  b \, ,\qquad b \rightarrow  -c^{(1)} \, ,\\
\end{equation}
and:
\begin{equation}
{\tilde c}^{(5)} \rightarrow  c^{(5)} \, ,
\qquad c^{(5)} \rightarrow  -{\tilde c}^{(5)} \, .\\
\end{equation} 

\vspace*{1cm}

Therefore we have found that the IIB NS-5-brane, defined as the $T$-dual of
the IIA KK-monopole, is $S$-dual to the D-5-brane, as it was  
known already at the level of the corresponding type IIB 
supergravity solutions. 
It is interesting to note that the BI field transforms into a 1-form, 
and not into a 3-form as one would have expected. See the conclusions
for a further discussion on this point.

The worldvolume fields present in the IIB NS-5-brane effective action
describe the soliton solutions constructed in \cite{Papadopoulos}, 
as we are now going to see. We find the same worldvolume solitons than
for the IIA KK-monopole, given that 
both branes are $T$-dual to each other. Also, they are related by $S$-duality
to the worldvolume solitons of the D-5-brane 
\cite{CM,Gibbons,HLW}. 

The worldvolume field $c^{(1)}$ couples to the 0-brane and the 2-brane 
solitons (to the latter through its worldvolume dual 3-form).
Some corresponding intersections are: $(0|D1,NS5)$ and 
$(2|D3,NS5)$. They are related by $T$-duality to the
IIA KK-monopole solitons described by the configurations:
$(0|D2,KK)$ and $(2|D4,KK)$, respectively.

There are also two 3-brane solitons corresponding to the intersections:
$(3|NS5,NS5)$ and $(3|NS5,KK)$. The first one is obtained by applying 
$T$-duality to either of the two intersections in IIA: $(3|NS5,KK)$ or
$(3|KK,KK)_1$ (here the two $S^1$'s of the monopoles coincide). The
second is obtained from the configuration $(3|KK,KK)_2$ in IIA. The 4-forms in
the worldvolume of the IIB NS-5-brane that couple to these solitons
are the duals of the embedding scalars in the $5+1$ dimensional worldvolume.

${\tilde c}^{(5)}$ describes the tension of the NS-5-brane, 
and it couples to the
domain wall in the six dimensional worldvolume given by the intersection:
$(4|D5,NS5)$ \cite{BVP}. This soliton is $T$-dual to the 4-brane soliton
$(4|D4,KK)$ of the IIA KK-monopole.


\section{The IIB KK-monopole}
\label{IIBKK}


In this section we derive the action of the IIB KK-monopole through a
$T$-duality transformation 
in the action of the IIA NS-5-brane. In this case $T$-duality is
performed along a coordinate transverse to the 5-brane.

\vskip 16pt
\begin{figure}[!ht]
\begin{center}
\leavevmode
\epsfxsize= 10cm
\epsffile{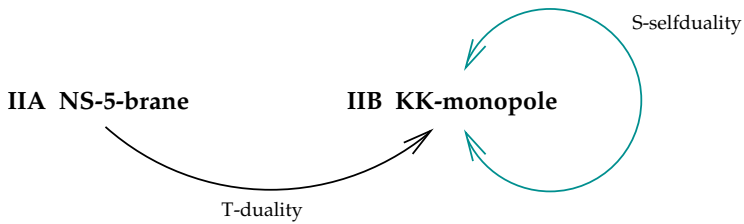}
\caption{\footnotesize
In this figure we depict how we have obtained the IIB KK-monopole action.
We apply $T$-duality on the IIA NS-$5$-brane action
along one coordinate in the transverse space.
One check of the action obtained is its invariance under $S$-duality.}
\label{fig:cuadro4}
\end{center}
\end{figure}

\subsection{The IIA NS-$5$ action}

Let us first recall the action of the 
type IIA NS-$5$-brane. In the quadratic approximation it is given by
\cite{BLO}
\begin{equation}
\begin{array}{rcl}
S &=&
-T_{\rm A5}\int d^6 \xi e^{-2 \phi} \sqrt{| {\rm det} \left(
g - (\alfa)^2 \ e^{2 \phi} {\cal G}^{(1)} {\cal G}^{(1)} \right)|} \times
\\& &\\& &
\times \left\{ 1 - {1 \over 4 \cdot 3!} (\alfa)^2 e^{2 \phi}
{\cal H}^{(3) \, 2} + \dots \right\}
\\& &\\& &
- (\alfa) {1 \over 6!} T_{\rm A5} \int d^6 \xi \epsilon^{i_1 \dots i_6}
{\tilde {\cal F}}^{(6)}_{i_1 \dots i_6} \, ,\\
\end{array}
\label{IIANS5}
\end{equation}

\noindent where ${\cal G}^{(1)}$ is defined as:

\begin{equation}
{\cal G}^{(1)} = \partial c^{(0)} + {1 \over \alfa} C^{(1)} \, ,
\label{GIIA}
\end{equation}

\noindent and the selfdual 3-form ${\cal H}^{(3)}$ as:

\begin{equation}
{\cal H}^{(3)}= 3 \partial a^{(2)} 
 + {1 \over \alfa} C^{(3)} + 3 \partial c^{(0)} B \, .\\
\end{equation}

The selfduality condition is inherited from that for the
M-5-brane, and takes the form:
\begin{equation}
{\cal H}^{(3)}_{ijk}= 
     \tfrac{1}{3!\sqrt{|\det ( g - (\alfa)^2 \ e^{2\phi}
                             {\cal G}^{(1)}{\cal G}^{(1)} ) | }}
                      \varepsilon_{ijklmn} {\cal H}^{(3)\ lmn}\, .
\label{selfdual5A}
\end{equation}

The 6-form ${\tilde {\cal F}}^{(6)}$ is the WZ-curvature
associated to the 5-form ${\tilde b}$, which describes the
tension of the IIA NS-$5$-brane:
\begin{equation}
\begin{array}{rcl}
{\tilde {\cal F}}^{(6)} &=&
6 \partial {\tilde b} + {1 \over \alfa}{\tilde B}
- 6 C^{(5)} \partial c^{(0)} + 30 C^{(3)} B \partial c^{(0)}
\\& &\\& &
- 30 \partial a^{(2)} C^{(3)} - 90 (\alfa) B \partial a^{(2)}
\partial c^{(0)} \, .\\
\end{array}
\end{equation}

The NS-5-brane contains a 1-brane, a 3-brane and a 5-brane solitons
\cite{Papadopoulos,BLO}. 
These are as well the soliton solutions occurring in
the M-5-brane effective action \cite{HLW,BGT}, given that both branes are
related by dimensional reduction.

The 1-brane, or self-dual string, couples to the
self-dual 2-form $a^{(2)}$, and it describes the boundary of a 
D-2-brane ending on the NS-5-brane: $(1|D2,NS5)$. The 3-brane
soliton couples to the dual of the worldvolume scalar $c^{(0)}$
and describes the intersection of two NS-5-branes: $(3|NS5,NS5)$.
Finally, if we considered the NS-5-brane on a massive background 
\cite{BLO} we would also find a coupling:
\begin{equation}
\int d^6\xi m\epsilon^{i_1\dots i_6}c^{(6)}_{i_1\dots i_6}\, ,
\end{equation}
which couples to a 5-brane 
soliton. This 5-brane soliton is the 
dimensional reduction of the M-5-brane 5-soliton,
realized as the embedding of the 
M-5-brane on an M-9-brane: $(5|M5,M9)$ 
\cite{deRoo}\footnote{The M-5-brane 5-brane soliton can be described
as well by the intersection \cite{Groningen-Boys,BGT}: 
$(5|M5,KK)$, though in this case
it is realized as a domain wall in the seven dimensional worldvolume of the 
eleven dimensional KK-monopole. This configuration gives the embedding
of the NS-5-brane on the KK-monopole: $(5|NS5,KK)$ after 
dimensional reduction.}.

The worldvolume solitons in the IIB NS-5-brane are related to these
solitons by a $T$-duality transformation along a worldvolume direction, 
as shown in \cite{Papadopoulos}. 

        It is important to notice that there is actually no compelling reason
why we should introduce the field ${\tilde b}$ describing the
tension of the IIA NS-$5$-brane. It has been pointed out
\cite{Townsend,BST} that since there is no $4$-brane soliton
on this brane, it is not necessary to elevate the brane tension
to the status of a dynamical variable. This can be  
argued from the fact that the IIA NS-$5$-brane is the dimensional
reduction of
the M-$5$-brane. The M-5-brane cannot have a boundary on any other brane
\cite{Townsend1},
so there is no need to replace its tension by a dynamical variable. 
{}From this point of view the role of ${\tilde b}$ in the IIA NS-5-brane
action is to guarantee invariance under all gauge transformations,
including total derivatives. It is also necessary in order to obtain the
worldvolume 5-form of the IIB NS-5-brane after a $T$-duality transformation
\cite{EJL}. This 5-form couples to the 4-brane soliton of the IIB NS-5-brane
\cite{BVP}.

The worldvolume field content is summarized in 
Table \ref{table-IIANS5} and the gauge 
symmetries of these fields are given in Appendix \ref{app-IIANS5}.

\begin{table}[t]
\begin{center}
\begin{tabular}{||c|c|c||}
\hline\hline
Worldvolume     & Field         & $\sharp$ of   \\      
Field           & Strength      & d.o.f         \\
\hline\hline
${X}^{{\mu}}$ & $-$ & $ 4 $ \\
\hline
$ c^{(0)}$ & $ {\cal G}^{(1)} $& $1 $\\
\hline
$ a^{(2)}_{ij}$& ${\cal H}^{(3)}$ & $3$\\
\hline
${\tilde b}_{i_1 \dots i_5}$ & ${\tilde {\cal F}}^{(6)}$
& $-$ \\
\hline\hline
\end{tabular}
\end{center}  
\caption{\label{table-IIANS5} {\bf Worldvolume fields of the IIA NS-5-brane}.
{\footnotesize In the case of the IIA NS-$5$-brane,
there are four embedding coordinates $X^\mu$, a scalar $c^{(0)}$
and a selfdual 2-form $a^{(2)}$. 
The 5-form ${\tilde b}$ describes the tension of the 5-brane.}}
\end{table}
 
\subsection{$T$-duality}

We apply now a $T$-duality transformation along a 
transverse direction.
The worldvolume fields do not change rank after $T$-duality since
we keep the same number of worldvolume dimensions. Moreover,
we have a new scalar field  $Z^\prime$, which is
the $T$-dual of the coordinate along which we perform 
the duality transformation.
The $T$-duality rules for the worldvolume fields are given by:
\begin{equation}
\begin{array}{rcl}
Z^{\prime} &=& (\alfa)\omega^{(0)} \, ,\\
& &\\
c^{(0) \prime} &=& -  {\tilde \omega}^{(0)} \, ,\\
& &\\
a^{(2) \prime} &=& - \omega^{(2)}+ 2(\alfa)
Z \partial \omega^{(0)} \partial {\tilde \omega}^{(0)} \, ,\\
& &\\
{\tilde b}^{\prime} &=& \omega^{(5)} 
- 30 (\alfa)^2 Z \partial\omega^{(2)} \partial \omega^{(0)} 
\partial {\tilde \omega}^{(0)} \, .\\
\end{array}
\label{IIANS5-->IIBKK}
\end{equation}
Notice that the
transversal coordinate and the scalar $c^{(0)}$ transform differently
than for D-branes.
In that case the transversal
coordinate $T$-dualizes into one of the components of the BI field
in the dual $D$-brane. Here, since we do not change
the number of worldvolume dimensions, this coordinate
$T$-dualizes into a new scalar field\footnote{
Furthermore, the scalar field $c^{(0)}$ (describing the
tension of a D-0-brane) is mapped into another scalar:
${\tilde \omega}^{(0)}$, whereas in a D-brane duality mapping it is mapped 
into the doubly reduced component of the 1-form $c^{(1)}$: 
$c^{(0) \prime} = c^{(1)}_\sigma$.}.
The occurrence of $Z$, the Taub-NUT coordinate of the KK-monopole, in
the right hand side of the last two expressions above is required by gauge
invariance, and assures the gauged sigma-model structure
necessary to describe the KK-monopole.

The worldvolume curvatures transform as follows:
\begin{equation}
\begin{array}{rcl}
{\cal H}^{(3) \prime} &=& - {\cal K}^{(3)} \, ,\\
& &\\
{\tilde {\cal F}}^{(6) \prime} &=& {{\cal K}}^{(6)} \, ,\\
& &\\
{\cal G}^{(1) \prime} &=& - \left( {\tilde {\cal K}}^{(1)} 
+ C^{(0)} {\cal K}^{(1)} \right) \, ,\\
\end{array}
\end{equation}
where ${\cal K}^{(3)}$
is the curvature of $\omega^{(2)}$,
and  ${\cal K}^{(1)}$ and ${\tilde {\cal K}}^{(1)}$ 
are the curvatures of $\omega^{(0)}$ and 
${\tilde \omega}^{(0)}$,
respectively. These field strengths are defined below.


\subsection{The action of the IIB KK-monopole}


The result of the $T$-duality transformation above is, again in quadratic 
approximation:
\begin{equation}
\begin{array}{rcl}
\label{monopollo}
S&=&
-T_{\rm BKK} \int d^6 \xi \,\, e^{-2 \varphi}k^2 \sqrt{|{\rm det} \left(
DX^\mu DX^\nu \j_{\mu \nu} - (\alfa)^2 k^{-2} e^{\varphi}
{\cal K}^T M {\cal K} \right)|} \times
\\& &\\& &
\times \left\{1 - {e^{2 \varphi} \over 4 \cdot 3!} (\alfa)^2 k^{-2} 
\, {({\cal K}^{(3)})}^2 
+ \dots\right\} \\& &\\& &
- (\alfa) {1 \over 6!} T_{\rm BKK} \int d^6 \xi \epsilon^{i_1 \dots i_6}
{\tilde {\cal K}}^{(6)}_{i_1 \dots i_6} \, .\\
\end{array}
\end{equation}
Here ${\cal K}^T M {\cal K}$ is the $SL(2, {\rm R})$-invariant
combination:
\begin{equation}
{\cal K}^T M {\cal K} = \left( \begin{array}{cc}{\cal K}^{(1)} 
& {\tilde {\cal K}}^{(1)} \end{array}
\right)  e^\varphi \left( \begin{array}{cc}
e^{-2 \varphi} + C^{(0)\, 2} &  C^{(0)} \\
C^{(0)} & 1 \end{array} \right)
\left( \begin{array}{c} {\cal K}^{(1)} \\ {\tilde {\cal K}}^{(1)}
\end{array} \right)
\, ,
\end{equation}
with
\begin{equation}
{\cal K}^{(1)} = \partial \omega^{(0)} - {1 \over \alfa}(i_k {\cal B}) 
\, ,\qquad
{\tilde {\cal K}}^{(1)} = \partial {\tilde \omega}^{(0)} 
+ {1 \over \alfa}(i_k C^{(2)}) \, ,
\end{equation}
the gauge invariant curvatures of the worldvolume scalars
$\omega^{(0)}$ and ${\tilde \omega}^{(0)}$.
The covariant derivative is the usual one for KK-monopoles:
\begin{equation}
DX^\mu = \partial X^\mu + A k^\mu \, ,
\end{equation}
with $k^\mu$ the Killing vector describing the Taub-NUT isometry
and $A$ the dependent field:
\begin{equation}
A=k^{-2} \partial X^\mu k_\mu \, ,
\end{equation}
where $k^2= - k^\mu k^\nu \j_{\mu \nu}$.
The curvature of the two form $\omega^{(2)}$ is given by:
\begin{equation}
\begin{array}{rcl}
{{\cal K}}^{(3)} &=&
3 \partial \omega^{(2)} + {1 \over \alfa} (i_k C^{(4)})
+{6}(\alfa) A \partial \omega^{(0)}
\partial {\tilde \omega}^{(0)} \\& &\\& &
+ {3 \over 2(\alfa)} (DX^\mu DX^\nu {\cal B}_{\mu \nu})(i_k C^{(2)}) 
- {3 \over 2(\alfa)} (DX^{\mu} DX^{\nu} C^{(2)}_{\mu \nu}) (i_k {\cal B})
\\& &\\& &
+ {3} \partial {\tilde \omega}^{(0)} DX^\mu DX^\nu {\cal B}_{\mu \nu}
+ {3} \partial \omega^{(0)} DX^\mu DX^\nu C^{(2)}_{\mu \nu}
\, .\\
\end{array}
\end{equation}

\begin{table}[t]
\begin{center}
\begin{tabular}{||c|c|c||}
\hline\hline
Worldvolume     & Field         & $\sharp$ of   \\      
Field           & Strength      & d.o.f         \\
\hline\hline
${X}^{{\mu}}$ &$-$ & $10 - 6 - (1) = 3$ \\
\hline
$ \omega^{(0)} $ &
${\cal K}^{(1)}$& $1$\\
\hline
${\tilde \omega}^{(0)}$ & ${\tilde {\cal K}}^{(1)}$& $1$ \\
\hline
$\omega^{(2)}$ & ${\cal K}^{(3)}$ & $3$\\
\hline
${\tilde \omega}^{(5)}$ 
& ${\tilde {\cal K}}^{(6)}$ & $-$ \\
\hline\hline
\end{tabular}
\end{center}  
\caption{\label{table-IIBKK} 
{\bf Worldvolume field content of the IIB KK-monopole}.
{\footnotesize In the case of the IIB KK-monopole,
there are three embedding coordinates $X^\mu$, after gauge fixing the
Taub-NUT coordinate,  
two worldvolume scalars $\omega^{(0)}$, ${\tilde \omega}^{(0)}$,
constituting a doublet under $S$-duality, 
a self-dual 2-form $\omega^{(2)}$ ($S$-selfdual) and a
5-form ${\tilde \omega}^{(5)}$, describing the tension
of the KK-monopole and also $S$-selfdual.}}
\end{table}

\noindent Finally, ${\tilde {\cal K}}^{(6)}$ is the field strength of the
5-form ${\tilde \omega}^{(5)}$ describing the tension of the
IIB KK-monopole:
\begin{equation}
\begin{array}{rcl}
{\tilde {\cal K}}^{(6)} &=&
6\partial {\tilde \omega}^{(5)} + {1 \over \alfa} (i_k {\cal N})
+{6} (i_k {\tilde {\cal B}}) \partial \omega^{(0)}
-{6} (i_k C^{(6)}) \partial {\tilde \omega}^{(0)}
\\& &\\& &
+30 (DX \dots DXC^{(4)})(i_k C^{(2)})\partial \omega^{(0)}
+30 (DX \dots DXC^{(4)})(i_k {\cal B})\partial {\tilde \omega}^{(0)}
\\& &\\& &
+30 (i_kC^{(4)}) (DXDXC^{(2)})\partial \omega^{(0)}
+30 (i_kC^{(4)}) (DXDX {\cal B})\partial {\tilde \omega}^{(0)}
\\& &\\& &
+ 45 (DXDXC^{(2)})(DXDX{\cal B})(i_kC^{(2)})\partial \omega^{(0)}
\\& &\\& &
- 45 (DXDX {\cal B})(DXDX C^{(2)})(i_k {\cal B})\partial {\tilde \omega}^{(0)}
\\& &\\& &
-30 (\alfa)\partial \omega^{(2)}\left( {\cal K}^{(3)} 
-6(\alfa) A\partial \omega^{(0)} \partial {\tilde \omega}^{(0)} \right)
\\& &\\& &
+60 (\alfa)^2 {\cal K}^{(3)}A \partial \omega^{(0)} \partial 
{\tilde \omega}^{(0)}
-30 (\alfa) (DX \dots DX C^{(4)}) \partial \omega^{(0)} \partial 
{\tilde \omega}^{(0)}
\\& &\\& &
-360 (\alfa)^2 \partial \omega^{2} A \partial \omega^{(0)} \partial 
{\tilde \omega}^{(0)}
\, .\\
\end{array}
\end{equation}

This action is invariant under local transformations in the
worldvolume: $\delta X^\mu=-\sigma(\xi) k^\mu$, with  
$A$ playing the role of gauge field:
$\delta A=\partial\sigma$. This symmetry can be gauge
fixed by eliminating the coordinate adapted to it, which is the Taub-NUT
direction of the monopole. In this way we obtain the right number of degrees of
freedom \cite{BJO}.

The field content of the worldvolume theory  
is summarized in Table \ref{table-IIBKK}.
It consists on a selfdual 2-form and 5 scalars, which is the field content
of the $(2,0)$ vector supermultiplet \cite{Callan}.
The worldvolume symmetries can be found in Appendix C.4.

The selfduality of the 2-form $\omega^{(2)}$ in the linear
approximation takes the form:
\begin{equation}
\label{selfi2}
{\cal K}^{(3)}_{i_1 \dots i_3} = {1 \over \sqrt{|{\rm det}(\Pi^{(B)})|}}
\Pi^{(B)}_{i_1 j_1} \dots \Pi^{(B)}_{i_3 j_3}
\epsilon^{j_1 \dots j_6} {\cal K}^{(3)}_{j_4 j_5 j_6} \, ,
\label{KKB-Poincare}
\end{equation}
where we have introduced a shorthand notation for the gauged sigma-model
metric: $\Pi^{(B)}_{ij} = D_iX^\mu D_jX^\nu \j_{\mu \nu}$
\footnote{In the kinetic term of (\ref{monopollo}) there is an
inverse metric $\Pi^{(B) \, ij}$ such that
$\Pi^{(B) \, ij}\Pi^{(B)}_{jk} = \delta^i_k$. Although
the reduced metric $\Pi^{(B)}_{\mu \nu} = \j_{\mu \nu} + k^{-2} k_\mu k_\nu$
has no inverse, the pull-back of this 
$\Pi^{(B)}_{ij}=\partial_i X^\mu \partial_j X^\nu \Pi^{(B)}_{\mu \nu}$
has a well defined inverse $\Pi^{(B) \, ij}$ 
in the $5+1$ dimensional worldvolume.}.

Notice that
we have introduced a new field $i_k {\cal N}$. This field 
is defined from the $T$-duality rule for the field ${\tilde B}$
of type IIA:

\begin{equation}
{\tilde B}^{\prime}_{\mu_1 \dots \mu_6}= {\cal N}_{\mu_1 \dots \mu_6 z} \, .\\
\end{equation}

\noindent Its gauge transformation rule can be found 
in the appendix (formula (\ref{Ntorcida})),
and shows that $i_k {\cal N}$ is S-selfdual.
The complete action of the IIB KK-monopole is in fact invariant under
$S$-duality. This duality works on the worldvolume fields as follows:
\begin{equation}
\begin{array}{rclrcl}
\omega^{(2)} &\rightarrow& \omega^{(2)} \, ,\hspace{1.0cm}&
{\tilde \omega}^{(5)} &\rightarrow& {\tilde \omega}^{(5)} \, ,\\
& &\\
\omega^{(0)} &\rightarrow& {\tilde \omega}^{(0)} \, ,&
{\tilde \omega}^{(0)} &\rightarrow& - \omega^{(0)} \, ,\\
\end{array}
\ee

\noindent i.e. $\omega^{(0)}$ and 
${\tilde \omega}^{(0)}$ constitute a doublet under $S$-duality
whereas $\omega^{(2)}$ and ${\tilde \omega}^{(5)}$ are S-selfdual.

Intersections in Type IIB can be cast into representations of
$S$-duality. This has the consequence that the solitons occurring
in the branes fit as well into $S$-duality representations.
This picture emerges naturally in the worldvolume action of the monopole.

The IIB KK-monopole has the same worldvolume soliton solutions as the IIA
NS-5-brane, given that both branes are related by $T$-duality 
\cite{Papadopoulos}. We show below that the worldvolume field content 
that we find in the IIB KK-monopole effective action precisely describes the
couplings to these solitons. 

The field $\omega^{(2)}$ couples to a 1-brane soliton on the brane,
which can be des\-cribed by the intersection of a D-$3$-brane with
the KK-monopole: $(1|KK, D3)$. Here one of the worldvolume directions of the
D-3-brane is wrapped around the $S^1$ of the monopole. 
This configuration is $S$ selfdual.
        
We can also have a 3-brane soliton on the KK-monopole,
which can be obtained by two different intersections:
\begin{equation}
(3|KK,D5) \, ,\qquad (3|KK,NS5) \, ,
\end{equation}
where in both cases one direction of the brane is wrapped
around the $S^1$ of the monopole. These two configurations are $S$-dual to 
each other, with each 3-brane soliton described by one of the
0-forms forming an $SL(2,Z)$ doublet: $(\omega^{(0)},
{\tilde \omega}^{(0)})$.  

        We still have the three embedding scalars, which may play  
a role in the construction of 3-brane solitons.
Since they are inert under $S$-duality, they must couple to solitons
that appear in $S$-selfdual configurations of branes. These are 
the two possible intersections of 
two IIB KK-monopoles on a three brane: $(3|KK,KK)_{1,2}$, which we also
encountered in the IIA theory.

The $5$-brane soliton can be realized as the
intersection of the IIB KK-monopole with a D-$7$-brane. The 
KK-monopole is completely embedded in the D-$7$-brane, which lies transversal
to the $S^1$ direction.

There are other intersections over 1-, 3- or 5-branes for which one can 
find out to which
worldvolume field they should couple, in view of the S-symmetry of the
configuration. 
As for the IIA NS-5-brane, there are no intersections over a 4-brane.
Therefore the same argument
presented in the previous section
against a dynamical tension for the IIA NS-$5$-brane 
\cite{Townsend,BST} applies to the IIB KK-monopole. 


\section{$T$-duality between type II NS-5-branes}


In this section we present the $T$-duality rules between both type II
NS-5-branes. The duality is achieved in this case by means of a double
dimensional reduction. We restrict ourselves to the kinetic part of the
actions. 

In this and the next section hatted (unhatted) 
worldvolume directions are six (five) dimensional.

We split the worldvolume directions into
${\hat \imath}=(i,\sigma)$, where $\sigma$ is the worldvolume coordinate
which is identified with the target space coordinate in the reduction.

The $T$-duality rules of the worldvolume fields present in the IIA NS-5
brane effective action are:
 
\begin{equation}
\begin{array}{rcl}
c^{(0) \prime} &=& - c^{(1)}_\sigma \, ,\\
& &\\
a^{(2) \prime}_{i \sigma} &=& - c^{(1)}_i \, ,\\
& &\\
a^{(2) \prime}_{i j} &=& - {\tilde c}^{(3)}_{i j \sigma} \, ,\\
\end{array}
\end{equation}

\noindent where ${\tilde c}^{(3)}$ is the  worldvolume dual of  
$c^{(1)}$ \footnote{${\tilde c}^{(3)}$ is related to the 
3-form $c^{(3)}$
describing the tension of the 
explicitly $S$-selfdual D-$3$-brane \cite{CW}
by ${\tilde c}^{(3)} = c^{(3)} + (\alfa) c^{(1)} \partial b$.}. 
This duality can be seen as inherited
from the selfduality of $a^{(2)}$ in the IIA NS-$5$-brane,
and has the following
form in the linear approximation:

\begin{equation}
\label{selfi}
{\tilde {\cal G}}^{(4)}_{{\hat \imath}_1 \dots {\hat \imath}_4}
= {1 \over 2! \sqrt{|\j|} \sqrt{1 + e^{ 2 \varphi} C^{(0) \, 2}}}
\j_{{\hat \imath}_1 {\hat \jmath}_1} \dots \j_{{\hat \imath}_4 
{\hat \jmath}_4}
\epsilon^{{\hat \jmath}_1 \dots {\hat \jmath}_6} 
{\tilde {\cal F}}_{{\hat \jmath}_5 {\hat \jmath}_6} \, ,
\label{5B-Poincare}
\end{equation}

\noindent where ${\tilde {\cal G}}^{(4)}$ is the curvature of 
${\tilde c}^{(3)}$:

\begin{equation}
{\tilde {\cal G}}^{(4)} = 4 \partial {\tilde c}^{(3)} 
+ {1 \over \alfa}C^{(4)} + {3 \over \alfa}{\cal B} C^{(2)}
+ 12 {\cal B} \partial c^{(1)} \, .
\end{equation}

Using the $T$-duality transformation rules for the target space fields given in
Appendix A we find that the worldvolume field strengths transform as:

\begin{equation}
\begin{array}{rclrcl}
{\cal G}^{(1) \prime}_\sigma &=& 
              -\frac{1}{2\pi\alpha^\prime} C^{(0)} \, , \hspace{1.5cm} &
{\cal G}^{(1) \prime}_i &=&
              - {\tilde {\cal F}}_{i\sigma} \, , \\
& & & & \\
{\cal H}^{(3) \prime}_{ij\sigma} &=& 
              - {\tilde {\cal F}}_{ij} \, , &
{\cal H}^{(3) \prime}_{ijk} &=& 
               - {\tilde {\cal G}}^{(4)}_{ijk\sigma} \, .\\
\end{array}
\end{equation}

\noindent ${\tilde {\cal G}}^{(4)}_{ijk\sigma}$ is related to 
${\tilde {\cal F}}$ through (\ref{selfi}).

It is important to stress the fact that in the dualization of 
${\cal G}^{(1)}_\sigma$ a factor $2\pi\alpha^\prime$ enters in the
denominator. This implies that higher order terms in
$\alpha^\prime$ in the kinetic part of the IIA NS-5-brane will contribute
after duality to a lower order approximation for the IIB NS-5-brane.
 
The action (\ref{IIANS5}) of the IIA NS-5-brane is valid up to quadratic
order in $\alpha^\prime$. If we want to obtain an action for the IIB
NS-5-brane up to the same order we need to consider
the following, fourth order, expression for the
kinetic term of the IIA NS-5-brane:

\bea
S &=&
-T_{\rm A5}\int d^6 {\hat \xi} e^{-2 \phi} \sqrt{| {\rm det} \left(
g_{{\hat \imath}{\hat \jmath}} - 
(\alfa)^2 e^{2 \phi} {\cal G}^{(1)}_{\hat \imath} 
{\cal G}^{(1)}_{\hat \jmath} \right)|}
 \times 
                  \nn 
&& \times \left\{ 1 - \tfrac{(\alfa)^2}{4!} \ e^{2 \phi}\ 
({\cal H}^{(3)})^2 
                                          \right.        \nn 
          &&  \hspace{1.5cm} \left.    - \tfrac{3(\alfa)^4}{4!} 
                      \tfrac{e^{4\phi}}{1-(\alfa)e^{2\phi}{\cal G}^2}
           \cG^\hi \cG_\hj \cH_{\hi\hk\hl}\cH^{\hj\hk\hl}  +... 
       \right\} \, .
\eea
Note that this extra term does not come from considering the next order
contribution to the M-5-brane action,
 but from the inverse metric 
in the reduction of the $\cH^2$-term from 
eleven to ten dimensions.

$T$-duality in this action yields the action of the IIB NS-$5$-brane
with one extra 3-form ${\tilde c}^{(3)}$, i.e. 
the IIB NS-5-brane in a ``1- 3- form'' formulation.
We need to perform the
Poincar{\'e} duality (\ref{5B-Poincare}) in order to obtain the action 
(\ref{5brane-action}) in the ``1-form'' formulation. The result is, 
up to quadratic order in $\alpha^\prime$:

\bea
S&=& -T_{B5} 
\int d^6 {\hat \xi} e^{-2\varphi} \sqrt{1+e^{2\varphi}(C^{(0)})^2} \  
                                 \sqrt{|\det \j|} \x \nn
  && \times \left\{ 1 + \tfrac{(\alfa)^2}{4}\tfrac{e^{2\varphi}}
{1+e^{2\varphi}(C^{(0)})^2}
                      {\tilde {\cal F}}_{{\hat \imath}{\hat \jmath}} 
{\tilde {\cal F}}^{{\hat \imath}{\hat \jmath}}  +...  \right\} \ .
\eea

This provides a further check of the action that we have presented in
section 2.3 for the IIB NS-5-brane.


\section{$T$-duality between type II KK-monopoles}


We can do the same calculation as in the previous section
for the case of the type II KK-monopoles. Also here we concentrate only
on the kinetic terms.
The duality is achieved again by means of a double dimensional
reduction. The gauged isometry direction is kept the same in
both monopoles. 

The transformations of the worldvolume fields in the IIA 
KK-monopole are:
\begin{equation}
\begin{array}{rclrcl}
\omega^{(0) \prime} &=& \omega^{(0)} \, , \hspace{1.5cm} &
\omega^{(1) \prime}_\sigma &=& {\tilde \omega}^{(0)} \, ,\\
& & & &\\
\omega^{(1) \prime}_i  &=& \omega^{(2)}_{i \sigma}
\, , &
\omega^{(3) \prime}_{ij \sigma}  &=& \omega^{(2)}_{ij}
\, .\\
\end{array}
\end{equation}

The 2-form $\omega^{(2)}_{ij}$ in the IIB KK-monopole
is obtained from the $T$-dualization of a 
3-form, which is the worldvolume dual of the 1-form $\omega^{(1)}$
in the IIA KK-monopole action.  The selfduality
of the 2-form $\omega^{(2)}$ in the IIB KK-monopole translates into the
duality between $\omega^{(1)}$ and $\omega^{(3)}$.
Explicitly, the selfduality condition of $\omega^{(2)}$ 
(eq. (\ref{selfi2})), becomes after $T$-duality: 

\begin{equation}
{\cal K}^{(4)}_{{\hat \imath}_1 \dots {\hat \imath_4}} = 
{1 \over 2! \sqrt{{\rm det \Pi^{(A)}}} 
\sqrt{1+e^{2 \phi} k^{-2} (i_k C^{(1)})^2}}
\Pi^{(A)}_{{\hat \imath}_1 {\hat \jmath}_1} 
\dots \Pi^{(A)}_{{\hat \imath}_4 {\hat \jmath}_4} \epsilon^{{\hat \jmath}_1
\dots {\hat \jmath}_6} {\cal K}^{(2)}_{{\hat \jmath}_5 {\hat \jmath}_6}
\, ,
\label{KKA-Poincare}
\end{equation}
where ${\cal K}^{(4)}$ is the curvature of the 3-form $\omega^{(3)}$:
\begin{eqnarray}
{\cal K}^{(4)}& =& 4 \partial \omega^{(3)} + {1 \over \alfa}
(i_k C^{(5)}) + 4 (DXDXDXC^{(3)}) {\cal K}^{(1)} \nonumber\\
&&+12 (DXDXB) \partial \omega^{(1)}
+ 24 (\alfa) A \partial \omega^{(0)}
\partial \omega^{(1)} \, ,
\end{eqnarray}

\noindent and we have defined $\Pi^{(A)}_{{\hat \imath}{\hat \jmath}}=
D_{\hat \imath}X^\mu D_{\hat \jmath}X^\nu g_{\mu\nu}$.

The 3-form $\omega^{(3)}$ is therefore associated to the target space field 
$i_k C^{(5)}$ of the IIA KK-monopole, and it couples to its 2-brane soliton. 
See the conclusions for a further discussion on this point.

The worldvolume duality transformations above, 
together with the transformations 
for the target space fields that can be found in Appendix A, give the action 
of the IIB KK-monopole in quadratic approximation.
Note that in this calculation it is necessary to perform the Poincar{\'e}
duality transformation (\ref{KKA-Poincare}) in order to obtain the
selfduality condition (\ref{selfi2}) of the IIB KK-monopole action. 
  
This result provides a
further check of the action presented in section 3.3.


\section{Conclusions}


In this paper we have constructed the worldvolume effective actions of the
NS-5-brane and KK-monopole of the type IIB theory. 
Their worldvolume field content is
precisely what is needed in order to explain the soliton configurations
of these branes. 

It is remarkable that the worldvolume fields that couple to the soliton
solutions of a six dimensional brane are those that are needed to
construct invariant field strengths for each target space field coupled to
the brane. These field strengths take the form:
$p \partial c^{(p-1)}+\frac{1}{2\pi\alpha^\prime}C^{(p)}+\dots$.
Consider for instance the IIB NS-5-brane. 
The target space fields that couple to its worldvolume are:
$C^{(2)}, C^{(4)}$ and ${\cal B}$. The corresponding worldvolume
field strengths contain the worldvolume fields: $c^{(1)}, c^{(3)}$ and 
a 5-form ${\bar  c}^{(5)}$,
which couple to 0-brane, 2-brane and 4-brane solitons, respectively.
Given that $c^{(3)}$ is the worldvolume dual of $c^{(1)}$ it is possible
to construct an action in which only $c^{(1)}$ is present, which is the one 
that we have constructed in section 2. If $c^{(3)}$ is included then 
the 5-form that enters the field strength of ${\cal B}$ (the WZ term)
is ${\bar c}^{(5)}$, which
differs from ${\tilde c}^{(5)}$ by a worldvolume field 
redefinition\footnote{That is  why we have denoted it 
${\bar c}^{(5)}$ and not ${\tilde c}^{(5)}$, as it appears in the
IIB NS-$5$-brane action of section 2.}. 
One can conclude that a $(p-2)$-brane soliton in the worldvolume of a
$q$-brane couples to its worldvolume field $c^{(p-1)}$ and describes
the boundary of a $(p-1)$-brane ending on the $q$-brane.
The eleven
dimensional soliton configurations derived in \cite{BGT} can also be
associated to the worldvolume fields assigned to each target space field
occurring in the worldvolume.
This idea provides a very simple way of 
identifying which soliton solutions should be found in a particular
brane. 
That this could be the case was anticipated in \cite{Ricc,CW}.

For the KK-monopoles, the worldvolume fields are associated to
the contractions of the Killing vector with the target space fields.
The field strengths take the form:
$p\partial\omega^{(p-1)}+\frac{1}{2\pi\alpha^\prime}
(i_k C^{(p+1)})+\dots$. In this case 
$\omega^{(p-1)}$ couples to a $(p-2)$-brane soliton, which describes 
the boundary of a
$p$-brane ending on the KK-monopole, with one of its worldvolume
directions wrapped around the Taub-NUT direction of the monopole. 
Let us consider, for instance,
the case of the IIA KK-monopole. The target space field associated to
$\omega^{(1)}$ is $i_k C^{(3)}$.
Then $\omega^{(1)}$ describes a D-2-brane, wrapped around the Taub-NUT 
direction, ending on the KK-monopole. 

The mechanism by which we have constructed the previous actions is
$T$-duality. The corresponding objects in IIA which the NS-5-brane and the 
KK-monopole in IIB are dual to are the IIA KK-monopole and the IIA
NS-5-brane. These branes contain couplings to dual target space fields in
their worldvolumes, as well as to non-trivial worldvolume fields, for which
the $T$-duality transformation rules were not known. We have constructed the
$T$-duality rules for all the target space and worldvolume fields
coupled to these actions.

One possible generalization of
type IIA supergravity contains a cosmological constant
term. This is the so-called massive supergravity \cite{Romans,BRGPT}. This
theory is related to type IIB supergravity by massive $T$-duality rules,
which have been derived in the literature for the gauge  potentials
occurring in the supergravity actions, as well as for the worldvolume
fields coupled to
some low dimensional
D-branes \cite{BRGPT,Eric-Mees,Green-Hull-Townsend}. 
We have left for a future
publication the derivation of the complete set of $T$-duality rules that map
other solitonic objects, where dual gauge potentials and more general
worldvolume fields are also present in the effective actions \cite{EJL}.

An interesting feature about the IIB NS-5-brane is the way
the worldvolume fields must transform under $S$-duality in order to show its
connection with the D-5-brane. One would have expected the
corresponding actions to be
related by an $S$-duality transformation for the target space fields plus 
an additional worldvolume duality for the BI field, with the
consequence that the NS-5-brane would have depended on a worldvolume
3-form \cite{Hull}. This is in fact the way the F-string and the D-string 
are related \cite{D2-brane}. There, together with
the $S$-duality transformation of the backgrounds, one needs to perform a
worldvolume duality transformation giving rise to the Born-Infeld field of
the D-string. 
However the worldvolume field content of the NS-5-brane reveals that this
Poincar{\'e} duality does not take place. Instead, we have found that the 
$S$-duality
transformation rules of the worldvolume fields reveal that they
transform as doublets or singlets depending on the behaviour under
$S$-duality of the target space field to which they are associated.
We believe this information may be relevant for the construction of the
worldvolume action of the so-called $(p,q)$ 5-brane multiplet 
\cite{Hull,Witten1}\footnote{Recently a $(p,q)$ multiplet of 5-brane
  solutions of type IIB supergravity has been constructed in \cite{roy}.}.

Moreover, this result suggests that type IIB solitonic branes
which are $S$-dual partners of D-$p$-branes, 
i.e. IIB (1,0) $p$-branes,
with $p$ odd and $p>3$,
are described by an effective action whose kinetic term contains
the 1-form $c^{(1)}$, and is given by:
\begin{equation}
\int d^{p+1}  \xi \, e^{- ( {p-1 \over 2})\varphi }
( 1 + e^{2 \varphi} C^{(0) \, 2})^{ p-3 \over 4}
\sqrt{| {\rm det} \left( \j + { (\alfa) e^\varphi \over
\sqrt{1+e^{2\varphi}C^{(0) \, 2}}} {\tilde {\cal F}} \right)|} \, .
\label{hartura}
\end{equation}
This is the $S$-dual of the kinetic term of the corresponding 
D-$p$-brane. The $p=5$ case corresponds to the 5-brane that we have 
obtained in this article.
For $p=7$ (\ref{hartura}) describes the kinetic term of a 
(1,0) 7-brane, which can in
fact be obtained as  $T$-dual to the
M-KK-monopole of \cite{BEL} reduced
along a transversal coordinate \cite{EJL}. 
This information can be relevant for the construction of the
worldvolume actions of other type IIB $(p,q)$ brane multiplets
\footnote{See \cite{PaTo} for some recent work in this direction.}.  

We have found that the type IIB KK-monopole is described by a 
gauged sigma model, such that
the degree of freedom associated to its Taub-NUT isometry is effectively
eliminated from the action. Gauged sigma-models have proved very useful
in order to describe Kaluza-Klein monopoles and eleven dimensional 
massive branes \cite{BLO,LO}.

It was shown in \cite{BLO} that the definition of a massive eleven
dimensional supergravity from which Romans' massive IIA SUGRA could be
derived upon dimensional reduction, requires the existence of a Killing
isometry in the eleven dimensional background. Massive IIA D-branes are
obtained through direct and double dimensional reduction from massive
M-branes, whose effective actions are described by gauged sigma-models in
which the Killing isometry is gauged.
The M-9-brane is a solution of
massive eleven dimensional supergravity, since its double dimensional
reduction gives the massive IIA D-8-brane \cite{PW,BRGPT}. It has been
shown recently\footnote{For the gravitational part.} \cite{BvdS} 
that the M-$9$-brane is described by the same kind of gauged
sigma-model as the KK-monopoles.
The IIA NS-9-brane obtained by reduction along the transverse
coordinate contains the correct dilaton coupling \cite{Hull}. Also,
the IIB D-9-brane obtained from it after T and S dualities has the correct
scaling with $e^{-\phi}$. In this paper we have obtained explicit T and S 
duality 
rules for worldvolume fields that couple as well to the worldvolume actions 
of 9-branes. For instance, we know from $S$-duality in the IIB D-9-brane that 
the IIB NS-9-brane must contain the worldvolume vector $c^{(1)}$ in its
effective action.
Many of the results for 5-branes in this article can be generalized to
9-branes. We hope to report progress in this direction in the near future.


\section*{Acknowledgements}
We would like to thank E.~Bergshoeff and T.~Ort\'{\i}n
 for useful discussions.
E.E.~would also like to thank R.~Argurio, M.~de Roo and  
J.P.~van der Schaar for discussions.
The work of E.E.~and B.J.~is part of the research
program of the "Stichting FOM".
The work of Y.L.~is supported by a TMR fellowship from the
European Commission.


\appendix


\section{$T$-duality}
\label{T-duality}


In this appendix we give the $T$-duality rules for the 
background fields that couple to the NS-5-branes and KK-monopoles.
For completion we also summarize the target space $T$-duality rules that
have been constructed previously in the literature \cite{BHO,
  Green-Hull-Townsend}. In our notation 
$z$ is the direction along which we perform the duality transformation.

The $T$-duality rules that map $(i_k N)$ and
$(i_k {\tilde B})$, coupled to the IIA KK-monopole action, onto type
IIB backgrounds are:

\begin{equation}
\begin{array}{rcl}
{N}^{\prime}_{\mu_1 \dots \mu_6 z} &=&
- {\tilde {\cal B}}_{\mu_1 \dots \mu_6} + 
{6}{\tilde {\cal B}}_{[\mu_1 \dots \mu_5 z}{\j_{\mu_6] z} \over \j_{zz}} +
{20}C^{(4)}_{[\mu_1 \dots \mu_3 z} C^{(2)}_{\mu_4 \mu_5}
{\j_{\mu_6] z} \over \j_{zz}} \\
& & \\
& &
-{15 \over 2} {\cal B}_{[\mu_1 \mu_2} C^{(2)}_{\mu_3 \mu_4}
C^{(2)}_{\mu_5 \mu_6]}
- {15}{\cal B}_{[\mu_1 z} C^{(2)}_{\mu_2 \mu_3}
C^{(2)}_{\mu_4 \mu_5} { \j_{\mu_6] z} \over \j_{zz}}\\
& & \\
& &
+ 60 {\cal B}_{[\mu_1 \mu_2} C^{(2)}_{\mu_3 \mu_4}
C^{(2)}_{\mu_5 z} {\j_{\mu_6] z} \over  \j_{zz}}
\, ,
\\
& &
\\
{\tilde B}^{\prime}_{\mu_1 \dots \mu_5 z} &=&
{\tilde {\cal B}}_{\mu_1 \dots \mu_5 z} + 5 C^{(4)}_{[\mu_1 \dots \mu_4}
C^{(2)}_{\mu_5] z} + 5 C^{(4)}_{[\mu_1 \dots \mu_3 z}C^{(2)}_{\mu_4 \mu_5]}
+{10}C^{(4)}_{[\mu_1 \dots \mu_3 z} C^{(2)}_{\mu_4 z}
{\j_{\mu_5 ] z}  \over \j_{zz}}\\
& &\\& &
{15 \over 2}{\cal B}_{[\mu_1 \mu_2} C^{(2)}_{\mu_3 \mu_4} C^{(2)}_{\mu_5] z}
+ 15 {\cal B}_{[\mu_1 z} C^{(2)}_{\mu_2 z} C^{(2)}_{\mu_3 \mu_4}
{\j_{\mu_5] z} \over \j_{zz}} \, .\\
\end{array}
\end{equation}

\noindent {}From type IIB to type IIA we have:

\begin{equation}
\label{atres}
\begin{array}{rcl}
{\tilde {\cal B}}^{\prime}_{\mu_1 \dots \mu_5 z} &=&
{\tilde B}_{\mu_1 \dots \mu_5 z} 
-  5 \left( C^{(5)}_{[\mu_1 \dots \mu_4 z}
-3B_{[\mu_1 \mu_2} C^{(3)}_{\mu_3 \mu_4 z}\right)
\left(C^{(1)}_{\mu_5]} - C^{(1)}_z {g_{\mu_5] z} \over g_{zz}} \right) \\
& &\\& & 
-5 \left(C^{(3)}_{[\mu_1 \mu_2 \mu_3} 
- {3 \over 2} C^{(3)}_{[\mu_1 \mu_2 z}
{g_{\mu_3] z} \over g_{zz}}\right)
C^{(3)}_{\mu_4 \mu_5 z} 
\, ,
\\
& &
\\
{\tilde {\cal B}}^{\prime}_{\mu_1 \dots \mu_6} &=& 
- N_{\mu_1 \dots \mu_6 z} 
-6 {\tilde B}_{[\mu_1 \dots \mu_5 z}B_{\mu_6] z} 
\\& &\\& &
+30\left(C^{(5)}_{[\mu_1 \dots \mu_4 z} 
-3 C^{(3)}_{[\mu_1 \mu_2 z}B_{\mu_3 \mu_4}\right)
\left(C^{(1)}_{\mu_5} - C^{(1)}_z 
{g_{\mu_5 z} \over g_{zz}} \right) B_{\mu_6] z}\\
& &\\& &
+10 \left(C^{(3)}_{[\mu_1 \dots \mu_3} 
-{3 \over 2} {g_{[\mu_1 z} \over g_{zz}} C^{(3)}_{\mu_2 \mu_3 z} \right)
C^{(3)}_{\mu_4 \mu_5 z} B_{\mu_6] z}
\\& &\\& &
-30 C^{(3)}_{[\mu_1 \mu_2 z}C^{(3)}_{\mu_3 \mu_4 z} 
{g_{\mu_5 z} \over g_{zz}}B_{\mu_6] z}
- {15 \over 2} C^{(3)}_{[\mu_1 \mu_2 z} C^{(3)}_{\mu_3 \mu_4 z} 
B_{\mu_5 \mu_6}\, .\\
\end{array}
\end{equation}

Now we summarize the $T$-duality rules that have been constructed previously
in \cite{BHO,Green-Hull-Townsend}. 

{}From IIA to IIB we have:
 
\begin{displaymath}
\begin{array}{rcl}
C^{(1) \prime}_z &=& - C^{(0)}  \, ,\\
& &\\
C^{(1) \prime}_\mu &=& - C^{(2)}_{\mu z} + C^{(0)} {\cal B}_{\mu z} \, ,\\
& &\\
C^{(3) \prime}_{\mu \nu z} &=& - C^{(2)}_{\mu \nu} +
{2} C^{(2)}_{[\mu z} {\j_{\nu ] z} \over \j_{zz}} \, ,\\
& &\\
C^{(3) \prime}_{\mu \nu \rho} &=& - C^{(4)}_{\mu \nu \rho z}
+{3 \over 2}C^{(2)}_{[\mu \nu}{\cal B}_{\rho] z}
\\& &\\& &
-{3 \over 2}{\cal B}_{[\mu \nu}C^{(2)}_{\rho] z}
\\& &\\& &
+ 6 C^{(2)}_{[\mu z}{\cal B}_{\nu z} {\j_{\rho] z} \over \j_{zz}}\, ,\\
\end{array}
\begin{array}{rcl}
e^{\phi^{\prime}} &=& {1 \over \sqrt{|\j_{zz}|}}  e^{\varphi} \, ,\\
& &\\
g^{\prime}_{\mu \nu} &=& \j_{\mu \nu} - {1 \over \j_{zz}}
\left(\j_{\mu z} \j_{\nu z} - {\cal B}_{\mu z} {\cal B}_{\nu z} \right)
\, ,\\
& &\\
g^{\prime}_{\mu z} &=& - {1 \over \j_{zz}} {\cal B}_{\mu z} \, ,\\
& & \\
g^{\prime}_{zz} &=& {1 \over g_{zz}} \, ,\\
& & \\
B^{\prime}_{\mu \nu} &=& {\cal B}_{\mu \nu} 
- {2 \over {\j}_{zz}}{\cal B}_{[\mu z} \j_{\nu] z} \, ,\\
& &\\
B^{\prime}_{\mu z} &=& - { \j_{z \mu} \over \j_{zz}} \, ,\\
\end{array}
\end{displaymath}
\begin{equation}
\begin{array}{rcl}
C^{(5) \prime}_{\mu_1 \dots \mu_4 z} &=& - C^{(4)}_{\mu_1 \dots \mu_4}
+ 4  C^{(4)}_{[\mu_1 \mu_2 \mu_3 z} {\j_{\mu_4] z} \over \j_{zz}}
- 3 C^{(2)}_{[\mu_1 \mu_2} {\cal B}_{\mu_3 \mu_4]}
\\& &\\& &
-6C^{(2)}_{z [\mu_1} {\cal B}_{\mu_2 \mu_3} {\j_{\mu_4] z} \over \j_{zz}}
-6{\cal B}_{z [\mu_1}C^{(2)}_{\mu_2 \mu_3} {\j_{\mu_4] z} \over \j_{zz}}
\, ,\\
& &\\
C^{(5) \prime}_{\mu_1 \dots \mu_5} &=& -C^{(6)}_{\mu_1 \dots \mu_5 z}
+5 \left( C^{(4)}_{[\mu_1 \dots \mu_4}
- 4 C^{(4)}_{[\mu_1 \dots \mu_3 z} 
{\j_{\mu_4 z} \over \j_{zz}}\right){\cal B}_{\mu_5]z}
\\& &\\& &
-{15 \over 2}{\cal B}_{[\mu_1 \mu_2}
{\cal B}_{\mu_3 \mu_4]}C^{(2)}_{\mu_5] z} 
- 30 C^{(2)}_{z [\mu_1} {\cal B}_{\mu_2 \mu_3}{\cal B}_{\mu_4 z}
{\j_{\mu_5] z} \over \j_{zz}}\, .\\ 
\end{array}
\end{equation}

\noindent Similarly, $T$-duality maps the type IIB background onto
the type IIA as follows: 
\begin{equation}
\begin{array}{rcl}
C^{(0) \prime} &=& - C^{(1)}_z  \, ,\\
& &\\
C^{(2) \prime}_{\mu z} &=& -C^{(1)}_\mu + 
C^{(1)}_z { g_{\mu z} \over g_{zz}}\, ,\\
& &\\
C^{(2) \prime}_{\mu \nu} &=& - C^{(3)}_{\mu \nu z} + 2C^{(1)}_{[\mu} B_{\nu] z}
\\& &\\& &- {2}C^{(1)}_z {g_{z [\mu} \over g_{zz}}B_{\nu] z} \, ,\\
& &\\
C^{(4) \prime}_{\mu \nu \rho z} &=& - C^{(3)}_{\mu \nu \rho}
+{3 \over 2} C^{(3)}_{[\mu \nu z}{g_{\rho] z} \over g_{zz}} 
\\& &\\& &
+ {3 \over 2} \left( C^{(1)}_{[\mu}-C^{(1)}_z {g_{[\mu z} \over g_{zz}}
\right) B_{\nu \rho]}\, ,\\
\end{array}
\begin{array}{rcl}
e^{\varphi^{\prime}} &=& {1 \over \sqrt{|g_{zz}|}} e^{\phi} \, ,\\
& &\\
{\cal B}^{\prime}_{\mu \nu} &=&
 B_{\mu \nu} - {2 \over g_{zz}} B_{[\mu z}g_{\nu] z} 
\, ,\\ & &\\
{\cal B}^{\prime}_{\mu z} &=& -{g_{\mu z} \over g_{zz}} \, ,\\
& &\\
\j^{\prime}_{\mu \nu} &=& g_{\mu \nu} - {1 \over g_{zz}} \left(
g_{\mu z} g_{\nu z} - B_{\mu z} B_{\nu z} \right) \, ,\\
& &\\
\j^{\prime}_{\mu z} &=&  - {1 \over g_{zz}}B_{\mu z} \, ,\\
& &\\
\j^{\prime}_{zz} &=& {1 \over g_{zz}} \, ,\\
\end{array}
\label{IIB-IIA-massive}
\end{equation}
\begin{displaymath}
\begin{array}{rcl}
C^{(4) \prime}_{\mu_1 \dots \mu_4} &=&
-C^{(5)}_{\mu_1 \dots \mu_4 z}
+4 \left( C^{(3)}_{[\mu_1 \mu_2 \mu_3}
-3 C^{(3)}_{[\mu_1 \mu_2 z} {g_{\mu_3} \over g_{zz}} \right)B_{\mu_4] z} 
\\& &\\& &
+3 C^{(3)}_{[\mu_1 \mu_2 z} \left(
B_{\mu_3 \mu_4]} -2 B_{\mu_3 z} {g_{\mu_4] z} \over g_{zz}} \right)
-6 \left( C^{(1)}_{[\mu_1} - C^{(1)}_z {g_{[\mu_1 z} \over g_{zz}}
\right) B_{\mu_2 \mu_3} B_{\mu_4] z}\, ,\\
& &\\
C^{(6) \prime}_{\mu_1 \dots \mu_5 z} &=&
-C^{(5)}_{\mu_1 \dots \mu_5}
+5 C^{(5)}_{[\mu_1 \dots \mu_4 z}{g_{\mu_5] z} \over g_{zz}}
\\& &\\& &
-15 C^{(3)}_{[\mu_1 \mu_2 z} B_{\mu_3 \mu_4} {g_{\mu_5] z} \over g_{zz}}
\\& &\\& &
+ {15 \over 2} \left(
C^{(1)}_{[\mu_1} - C^{(1)}_z {g_{[\mu_1 z} \over g_{zz}}
\right)
B_{\mu_2 \mu_3}B_{\mu_4 \mu_5]} \, .\\
\end{array}
\end{displaymath}


\section{Target Space Gauge Symmetries}


In this appendix we give the gauge symmetries for the
background fields that couple to the six dimensional worldvolume theories
considered in this paper.

\begin{itemize}

\item Type IIA:

\begin{equation}
\begin{array}{rcl}
\delta C^{(1)} & = & \partial \Lambda^{(0)}\, ,\\
& & \\
\delta B & = & 2\partial\Lambda\, ,\\
& & \\
\delta C^{(3)} & = & 3 \partial\Lambda^{(2)} +
3\partial\Lambda^{(0)} B\, ,\\
& & \\
\delta C^{(5)} & = & 5\partial \Lambda^{(4)} +30\partial 
\Lambda^{(2)}B 
+15\partial\Lambda^{(0)} BB\, ,\\
& & \\
\delta\tilde{B} & = & 6\partial\tilde{\Lambda} 
-30 \partial\Lambda^{(2)} C^{(3)}
+6\partial\Lambda^{(0)} \left( C^{(5)} -5C^{(3)}B \right)\, .\\
\end{array}
\end{equation}

The KK-monopole couples as well to a new field $(i_k N)$, with
transformation rule:

\begin{equation}
\begin{array}{rcl}
\delta (i_k N)
 &=& 6 \partial (i_k \Sigma^{(6)})
+60 (i_k C^{(3)}) \partial
 (i_k \Lambda^{(4)})
- 30 \partial (i_k {\tilde \Lambda})(i_k B)
\\ & & \\ & &
- 60 (i_k C^{(3)})(i_k C^{(3)})\partial \Lambda  
+ 120 \partial \Lambda^{(2)}(i_k C^{(3)}) (i_k B) \\
& & \\
& &
 +60 (i_k C^{(3)}) \partial
( i_k \Lambda^{(2)}) B 
- 40 C^{(3)} \partial
 (i_k \Lambda^{(2)}) (i_k B) \\
& & \\
& &
- 20 C^{(3)} (i_k C^{(3)})
\partial (i_k \Lambda) 
- \sigma^{(0)} k^\lambda \partial_\lambda (i_k N)
\, .\\
\end{array}
\end{equation}

\noindent When coupled to the KK-monopole all the target space fields 
transform as well with respect to its Killing isometry, as indicated 
by the last term in $\delta (i_k N)$.

\item Type IIB: We work in the basis of fields where
the 4-form $C^{(4)}$ is $S$-selfdual.
The gauge transformations are given by:
\begin{equation}
\begin{array}{rcl}
\delta {\cal B} &=& 2 \partial \Lambda \, ,
\\& &\\
\delta C^{(2)} &=& 2 \partial \Lambda^{(1)} \, ,
\\& &\\
\delta C^{(4)} &=& 4 \partial \Lambda^{(3)} + 6 \partial \Lambda^{(1)}
{\cal B} - 6 C^{(2)} \partial \Lambda \, ,\\
& &\\
\delta C^{(6)} &=& 6 \partial \Lambda^{(5)} + 60 \partial \Lambda^{(3)}
{\cal B} 
\\& &\\& &
+ 45 \partial \Lambda^{(1)} {\cal B}{\cal B}
-90 C^{(2)} {\cal B} \partial \Lambda \, ,\\
& &\\
\delta {\tilde {\cal B}} &=&
6 \partial {\tilde \Lambda}^{(5)} - 60 \partial \Lambda^{(3)} C^{(2)}
\\& &\\& &
+ 45  \partial \Lambda C^{(2)} C^{(2)}
-90 \partial \Lambda^{(1)} {\cal B} C^{(2)}\, .\\
\end{array}
\end{equation}

We also  include here the gauge transformation of the 
S-selfdual field $i_k {\cal N}$ which couples to the KK-monopole:

\begin{equation}
\label{Ntorcida}
\begin{array}{rcl}
\delta (i_k {\cal N})&=& 6 \partial ( i_k {\tilde \Sigma^{(6)}}) 
-30 \partial (i_k \Lambda^{(3)})(i_k C^{(4)})
\\& &\\& &
-6 ( i_k {\tilde {\cal B}}) \partial (i_k \Lambda)
- 6 (i_k C^{(6)}) \partial (i_k \Lambda^{(1)}) 
\\& &\\& &
-45 {\cal B}C^{(2)} (i_k C^{(2)}) \partial (i_k \Lambda)
-45 {\cal B}C^{(2)} (i_k {\cal B}) \partial (i_k \Lambda^{(1)})
\\& &\\& &
-30 C^{(4)} (i_k C^{(2)}) \partial (i_k \Lambda)
+30 C^{(4)} (i_k {\cal B}) \partial (i_k \Lambda^{(1)})
\\& &\\& &
-30 (i_k C^{(4)}) C^{(2)} \partial (i_k \Lambda)
+30 (i_k C^{(4)}) {\cal B} \partial (i_k \Lambda^{(1)})
\\& &\\& &
-45 \partial (i_k \Lambda^{(3)})(i_k C^{(2)}) {\cal B}
+45 \partial (i_k \Lambda^{(3)})(i_k {\cal B}) C^{(2)} \, .\\
\end{array}
\end{equation}

As for the IIA case all the fields transform as well with respect to the
Killing isometry of the IIB KK-monopole.

\end{itemize}


\section{Worldvolume Gauge Symmetries}


Here we give the gauge transformations of the worldvolume fields present
in the different actions that appear in this paper.


\subsection{IIA KK-monopole}


The gauge transformations of the worldvolume fields present in this action 
are \cite{BEL}:

\begin{equation}
\begin{array}{rcl}
\delta \omega^{(0)} &=& \frac{1}{2\pi\alpha^\prime}
(i_k \Lambda) \, ,\\ 
& & \\
\delta {\omega}^{(1)}&=&\partial \mu^{(0)}-
\frac{1}{2\pi\alpha^\prime}
(i_k \Lambda^{(2)})+
\Lambda^{(0)}\partial \omega^{(0)}
\, , \\ 
& & \\
\delta \omega^{(3)} &=& 3 \partial \mu^{(2)} -
{1 \over \alfa} (i_k \Lambda^{(4)}) - 3 \Lambda^{(2)} \partial \omega^{(0)}
\\& &\\& &
- 6 \Lambda \partial \omega^{(1)} - 6 (\alfa) \sigma^{(0)}
\partial \omega^{(0)} \partial \omega^{(1)} \, ,\\
& & \\
\delta \omega^{(5)}&=&5\partial\mu^{(4)}-
\frac{1}{2\pi\alpha^\prime}(i_k\Sigma^{(6)})
+5(i_k {\tilde \Lambda})\partial \omega^{(0)}+
20 (i_k\Lambda^{(4)}) \partial \omega^{(1)}
\\ & & \\
&&+60(2\pi\alpha^\prime)\Lambda^{(2)}\partial \omega^{(1)}
\partial\omega^{(0)}
+60(2\pi\alpha^\prime)\Lambda \partial \omega^{(1)}
\partial\omega^{(1)} \\ & & \\
&&+60(2\pi\alpha^\prime)^2\sigma^{(0)}
\partial\omega^{(1)}\partial \omega^{(1)} \partial\omega^{(0)} 
\, .
\end{array}
\end{equation}


\subsection{IIB NS-5-brane}


The worldvolume gauge transformations are given by:

\begin{equation}
\begin{array}{rcl}
\delta c^{(1)} &=& \partial \kappa^{(0)} - {1 \over \alfa} \Lambda^{(1)} \, ,\\
& &\\
\delta {\tilde c}^{(3)} &=& \partial \kappa^{(2)} 
- {1 \over \alfa} \Lambda^{(3)} - 6 \Lambda \partial c^{(1)} \, ,\\
& &\\
\delta {\tilde c}^{(5)} &=& 5 \partial {\tilde \kappa}^{(4)} +
{1 \over \alfa} {\tilde \Lambda}\\
& &\\
& & + 20 \Lambda^{(3)} \partial c^{(1)} + 60 (\alfa) \Lambda
\partial c^{(1)} \partial c^{(1)} \, .\\
\end{array}
\end{equation}


\subsection{D-$5$-brane}
\label{app-D5}


The worldvolume symmetries for the D-$5$-brane
are $S$-dual to those of the NS-$5$-brane:

\begin{equation}
\begin{array}{rcl}
\delta b &=& \partial \rho^{(0)} - {1 \over \alfa} \Lambda \, ,\\
& &\\
\delta {c}^{(5)} &=& 5 \partial {\kappa}^{(4)} -
{1 \over \alfa} {\Lambda}^{(5)}\\
& &\\
& & + 20 \Lambda^{(3)} \partial b - 60 (\alfa) \Lambda^{(1)}
\partial b \partial b \, .\\
\end{array}
\end{equation}


\subsection{IIA NS-$5$-brane}
\label{app-IIANS5}


The worldvolume symmetries are given by \cite{BLO}:

\begin{equation}
\begin{array}{rcl}
\delta c^{(0)} &=&  - {1 \over \alfa} \Lambda^{(0)} \, ,
\\& &\\
\delta a^{(2)} &=& 2\partial \rho^{(1)} - {1 \over \alfa}\Lambda^{(2)}
+ 2\partial c^{(0)} \Lambda \, ,
\\& &\\
\delta {\tilde b} &=& \delta {\tilde \rho}^{(4)}
-{1 \over \alfa}{\tilde \Lambda}
+ 5 \Lambda^{(4)} \partial c^{(0)} 
\\& &\\& &
- 15\Lambda^{(2)} \partial a^{(2)} - 30 (\alfa) \Lambda \partial a^{(2)}
\partial c^{(0)}\, .\\
\end{array}
\end{equation}


\subsection{IIB KK-monopole}

The worldvolume gauge transformations are:

\begin{equation}
\begin{array}{rcl}
\delta \omega^{(0)} &=& {1 \over \alfa}i_k \Lambda \, ,\\
& &\\
\delta {\tilde \omega}^{(0)} &=& - {1 \over \alfa} i_k \Lambda^{(1)} \, ,\\
& &\\
\delta \omega^{(2)} &=& 2\partial \mu^{(1)} - {1 \over \alfa}
i_k \Lambda^{(3)} - {2} \Lambda^{(1)} \partial \omega^{(0)}
- {2} \Lambda \partial {\tilde \omega}^{(0)} 
-{2}(\alfa)\sigma \partial \omega^{(0)} \partial {\tilde \omega}^{(0)} \, ,\\
& &\\
\delta {\tilde \omega}^{(5)} &=& 5\partial {\tilde \mu^{(4)}} 
-{1 \over \alfa}i_k {\tilde \Sigma}^{(6)} 
- {5} (i_k {\tilde \Lambda}) \partial \omega^{(0)}
+{5} (i_k \Lambda^{(5)}) \partial {\tilde \omega}^{(0)}
\\& &\\& &
+{20}(\alfa) \Lambda^{(3)} \partial \omega^{(0)} \partial {\tilde \omega}^{(0)}
- 15 (i_k \Lambda^{(3)}) \partial \omega^{(2)} 
\\& &\\& &
- 30 (\alfa) \Lambda^{(1)} \partial \omega^{(0)} \partial \omega^{(2)} 
- 30 (\alfa) \Lambda \partial {\tilde \omega}^{(0)} \partial \omega^{(2)}
\\& &\\& &
- 30 (\alfa)^2 \sigma \partial \omega^{(2)} 
\partial \omega^{(0)} \partial {\tilde \omega}^{(0)} \, .\\
\end{array}
\end{equation}


\end{document}